\documentclass[onecolumn,prd,superscriptaddress,nofootinbib,notitlepage]{revtex4-1}
\usepackage{latexsym}
\usepackage{amsmath,amsfonts,graphicx,accents}
\usepackage{amsbsy}
\usepackage{mathrsfs}
\usepackage{color}
\usepackage[colorlinks=true,
              linkcolor=blue,
              urlcolor=blue,
              citecolor=blue
              ]{hyperref}
\usepackage{psfrag}
\usepackage{enumerate}
\usepackage{amsmath,amssymb,calc,amsfonts}
\usepackage{latexsym}
\usepackage[utf8]{inputenc}
\usepackage[percent]{overpic}
\usepackage[autostyle=false, style=english]{csquotes}
\MakeOuterQuote{"}
\DeclareFontFamily{U}{rsfs}{}         
\DeclareFontShape{U}{rsfs}{m}{n}{<5> rsfs5 <6><7> rsfs7          %
  <8><9><10><10.95><12><14.4><17.28><20.74><24.88> rsfs10}{}     %
\DeclareMathAlphabet{\mathfs}{U}{rsfs}{m}{n}                     %
\definecolor{indiagreen}{rgb}{0.07, 0.53, 0.03}
\def\beq{\begin{equation}}
\def\eeq{\end{equation}}

\def\={\stackrel{\Delta}{=}}

\begin{document}
\title{Thermodynamics of Einstein-Gauss-Bonnet Black Holes under the Generalized Uncertainty Principle}

\author{Sasmita Kumari Pradhan} \email{sasmita.gita91@gmail.com}
\affiliation{Department of Physics, Centurion University of Technology and Management, Odisha, India}

 \author{Jamima Gee Varghese}\email{jamimageevarghese@gmail.com}
\affiliation{T. K. M. College of Arts and Science Kollam-691005, India}

\author{C. Fairoos}\email{fairoos.phy@gmail.com}
\affiliation{T. K. M. College of Arts and Science Kollam-691005, India}

\begin{abstract}

We explore the impact of the Generalized Uncertainty Principle(GUP) on the thermodynamics of five-dimensional Einstein-Gauss-Bonnet(EGB) black holes. A modified mass–temperature relation is derived under the assumption of local equilibrium, revealing that the black hole evolves into a stable remnant with a finite temperature, rather than undergoing complete evaporation. The modified entropy, obtained within this framework, deviates from the commonly expected logarithmic form. However, considering a linear GUP, which is obtained by combining Doubly Special Relativity (DSR) with the usual GUP, yields a logarithmic correction to the entropy similar to that of a five-dimensional Schwarzschild–Tangherlini black hole. However, the leading order term in the entropy of the EGB black hole is purely due to the underlying spacetime geometry, making the logarithmic term a higher order correction, unlike the case of the five-dimensional Schwarzschild black hole, where the leading order correction is logarithmic. Our results support the conjecture that the GUP-induced corrections to black hole entropy are sensitive to the dimensionality of spacetime.
\end{abstract}

 \maketitle
\section{Introduction}
Bekenstein's groundbreaking insight that a black hole possesses an entropy proportional to its horizon area signaled a profound and unexpected connection between black holes and the laws of thermodynamics \cite{Bekenstein:1972tm}. Subsequently, Hawking demonstrated that quantum field theory calculations near the event horizon predict that black holes emit thermal radiation, which would be detectable at infinity \cite{Hawking:1975vcx}. These ideas laid the foundation for the extensive study of black hole thermodynamics over the past several decades. Despite numerous advancements, the thermodynamic description of black holes remains a significant challenge. Notably, it has been demonstrated that black holes in asymptotically flat spacetimes exhibit negative heat capacity, resulting in an unstable, runaway radiation process that culminates in the well-known information loss paradox. Even though there are results that suggest the possible formation of a stable end state where the radiation process halts, potentially avoiding the information loss paradox. However, in the absence of a well-established quantum theory of gravity, a definitive resolution to this paradox remains elusive \cite{Susskind:1994vu,CALLAN1989673}.\\

Besides Hawking's original calculation, there are various methods to obtain the black hole temperature; see, for example, \cite{Parikh:1999mf}. For a spherically symmetric black hole, there exists a heuristic way to obtain the temperature by invoking the standard uncertainty principle \cite{ohanian1994gravitation}. Near the event horizon, there is a fundamental limit to how precisely we can know the position of a particle, roughly within the radius of the black hole event horizon. This positional fuzziness translates to an uncertainty in energy, which gives us an estimate of the energy (and thus temperature) of the particles being emitted. While heuristic arguments based on the uncertainty principle can provide an estimate of the characteristic energy scale of the emitted radiation, they do not, by themselves, explain why the spectrum of Hawking radiation should be thermal. The assumption of a blackbody spectrum arises primarily from the requirement of thermodynamic consistency, rather than from any direct derivation within the heuristic framework.\\

The need to go beyond the standard uncertainty principle arises naturally when gravity is incorporated into quantum theory. One of the most compelling consequences of this merger is the appearance of a fundamental minimal length scale, i.e., an observer-independent limit to how precisely spacetime distances can be resolved \cite{Kempf:1994su,Garay:1994en,Amelino-Camelia:2000cpa}. This notion of a minimal length is not speculative; it emerges in various well-established frameworks, including string theory \cite{AMATI198941}, non-commutative geometry \cite{Girelli:2004md}, and thought experiments involving micro-black holes at the Planck scale of spacetime \cite{Scardigli:1999jh}. The minimal length is typically expected to be of the order of the Planck length, suggesting a fundamental scale where classical concepts of spacetime break down. To consistently account for this minimal length within quantum mechanics, one is led to modify the Heisenberg uncertainty principle, resulting in what is known as the Generalized Uncertainty Principle (GUP) \cite{Adler:2001vs}. The corresponding mathematical expression reads,

\beq \label{GUP}
\delta x \delta p \geq \frac{\hbar}{2}\Bigg\{1+\frac{\beta^2 l_P^2}{\hbar^2}(\delta p)^2\Bigg\},
\eeq

where $l_P$ is the Planck length and $\beta$ is a dimensionless constant. The derivation of the GUP-modified Hawking temperature relies on identifying the black hole's horizon radius with the position uncertainty and the energy of the emitted particle with the momentum uncertainty. Applying the generalized uncertainty principle to these uncertainties yields the modified temperature. This heuristic approach, while simplified, effectively captures the leading-order quantum gravity corrections and is widely used in the literature. For instance, the introduction of GUP has profound implications for black hole thermodynamics, where it modifies the semiclassical relations such as the entropy-area law \cite{PhysRevD.70.124021} and leads to striking predictions like the existence of black hole remnants, i.e., stable, Planck-scale objects that could represent the final stage of black hole evaporation. Such features are also observed in alternative approaches, including Modified Dispersion Relations (MDR) \cite{Amelino-Camelia:2005zpp} and Doubly Special Relativity (DSR) \cite{Ali:2009zq}, pointing to a converging picture that suggests GUP may capture essential features of a quantum theory of gravity \cite{Nozari:2006ka, Nouicer:2007jg, Bina:2010ir, Banerjee:2010sd,Battista:2023iyu, Gialamas:2024ivq}.\\

Several significant works have applied the GUP to modify black hole thermodynamics, revealing important deviations from semiclassical predictions. In \cite{Majumder:2011xg}, author used the linear as well as quadratic form of GUP and computed corrections to the entropy of Schwarzschild and Reissner–Nordström black holes and found that the leading-order contributions include a positive term which is proportional to the square root of horizon area as well as a negative logarithmic term that precisely matches earlier results. It is found that the standard quadratic GUP yields a logarithmic entropy correction in six dimensions, achieving a logarithmic correction in five and seven dimensions requires either a linear GUP (as in DSR-inspired models) or adding cubic momentum terms \cite{Gangopadhyay:2015zma}. Interestingly, the GUP‑corrected entropy-area relation featuring a universal logarithmic term is shown to extend to a class of black objects such as Reissner–Nordström, Kerr, charged AdS, and spinning black rings \cite{Faizal:2014tea}. In \cite{Scardigli:2014qka}, the GUP Principle has been used to derive corrections to the Hawking temperature of Schwarzschild black holes, and in turn, to reconstruct a GUP-deformed Schwarzschild metric that consistently reproduces this modified temperature. Thus, the deformation parameter in GUP is directly associated with modifications to the spacetime geometry itself. Using this GUP-corrected metric, further analyses have revealed corrections to classical general relativistic predictions such as light deflection and perihelion precession, both in solar system dynamics and in binary pulsar systems. These observational effects have been employed to place upper bounds on the GUP deformation parameter, thus providing a phenomenological link between quantum gravity-inspired modifications and empirical data \cite{Das:2021lrb}.\\

In this paper, we investigate the thermodynamic modifications of five-dimensional Einstein–Gauss–Bonnet black holes arising from GUP. A central aim of this study is to examine the form of the GUP-induced corrections to the black hole entropy and, in particular, to test the conjectured universality of the leading-order logarithmic correction. EGB gravity offers a natural and well-motivated generalization of general relativity in higher dimensions, integrating the quadratic Gauss–Bonnet term into the action without introducing ghosts, and yielding second-order field equations \cite{Lovelock:1971yv}. This correction becomes non-trivial for dimensions greater than four, making five-dimensional spacetimes ideal for exploring non-Einstein effects. Black hole solutions in EGB gravity, such as those in asymptotically flat or (A)dS backgrounds, display modified thermodynamics, i.e., entropy no longer strictly follows the Bekenstein-Hawking area law, and phase structures including Hawking-Page transitions and thermodynamic stability depend sensitively on the Gauss–Bonnet coupling constant\cite{Cvetic:2001bk,Clunan:2004tb,Herscovich:2010vr}. These distinctive thermodynamic features, such as modified entropy, temperature, and heat capacity, make EGB black holes an ideal framework for investigating how GUP corrections interplay with higher-curvature effects, potentially leading to new insights.\\

The structure of the paper is as follows: In the next section, we briefly introduce the idea of GUP and discuss the thermodynamic modifications of Schwarzschild black holes. In Section \ref{three}, we incorporate GUP into the five-dimensional EGB black holes. Section \ref{four} is devoted to deriving the GUP as well as DSR-induced modifications to the entropy. Our main findings are summarized in Section \ref{dis}.

\section{General Uncertainty Principle and Thermodynamics of Schwarzschild Black Hole}\label{two}
 The Schwarzschild metric describes the spacetime outside a spherically symmetric, static, and chargeless mass distribution and is the simplest solution admitting a spacetime singularity portraying a black hole. The thermodynamic properties of such black holes are well-known \cite{Wald:1999vt}. It has been shown that when the GUP is considered, the spacetime structure as well as the thermodynamic behaviour of Schwarzschild black holes get modified\cite{Adler:2001vs,Majumder:2011xg, Gangopadhyay:2013ofa, Casadio:2020rsj}. In this section, we briefly outline the corrections to the black hole entropy due to the simplest form of the uncertainty principle given by Eq. \ref{GUP}. The metric function and event horizon($r_s$) for the Schwarzschild spacetime in general relativity are given by,

\beq
f(r) = 1-\frac{2 G M}{c^2 r}; \quad r_s=\frac{2 GM}{c^2}.
\eeq

Here, $M$ denotes the black hole mass, $G$ is Newton's gravitational constant in 4 dimensions, and $c$ is the speed of light. Let us consider a massless particle near the black hole horizon. The uncertainty in the position of the particle will be in the order of the Schwarzschild radius \cite{PhysRevD.70.124021,Nouicer:2007jg, KIM2007172,Alonso-Serrano:2020hpb}.

\begin{equation*}
    \delta x = 2\pi r_s,
\end{equation*}

 Assuming that the particle is in thermodynamic equilibrium with the hole, the momentum uncertainty near the horizon is given by,

\begin{equation*}
    \delta p = \frac{k_B T}{c}.
\end{equation*}

Here, $T$ is the black hole temperature and $k_B$ is the Boltzmann constant. Now, substituting $\delta x$ and $\delta p$ in Eq. \ref{GUP}, one finds the following expression for the GUP modified black hole mass as a function of temperature.

\beq\label{M_Sch}
M = \frac{M_P^2c^2}{8\pi}\Bigg\{\frac{1}{k_B T}+\beta^2 \frac{k_BT}{(M_Pc^2)^2}\Bigg\}.
\eeq

To arrive at the above expression, we have used the following relations between the Planck mass ($M_P$), Planck length ($l_P$), and $G$:

\beq
M_Pc^2=\frac{c\hbar}{l_P}; \quad M_P = \frac{c^2 l_P}{G}.
\eeq

The corresponding modification in temperature is captured in the following expression:

\beq \label{T_S}
T = \frac{4\pi Mc^2}{k_B \beta^2}\Bigg[1-\sqrt{1-\frac{\beta^2 M_P^2}{16 \pi^2 M^2}}\Bigg].
\eeq

As thermodynamic stability of a system is characterized by its heat capacity, we calculate the GUP modified heat capacity of the Schwarzschild black hole as,

\beq \label{H_C_S}
C=c^2\frac{dM}{dT}= \frac{k_B}{8\pi}\Bigg[(\beta^2-\left(\frac{M_Pc^2}{k_B T}\right)^2\Bigg].
\eeq

Note that the standard Schwarzschild black hole possesses a negative heat capacity, rendering it thermodynamically unstable. This instability manifests as a runaway process, i.e., as the black hole radiates, it loses mass, which in turn causes its temperature to rise. The increase in temperature accelerates the radiation further, eventually leading to the complete evaporation of the black hole. However, as shown in Eq.~\ref{H_C_S}, the introduction of the GUP modifies this behavior significantly. At low temperatures, specifically when \( k_B T \ll M_P c^2 \), the heat capacity remains negative, and the black hole continues to radiate. But as the mass decreases and the temperature increases, the GUP-corrected heat capacity becomes less negative and eventually reaches zero at a certain critical temperature. This critical point corresponds to the maximum attainable temperature of the black hole. Beyond this point, further evaporation is halted. The black hole stabilizes into a remnant with finite mass and non-zero temperature. An expression for the remnant mass can be easily obtained to be \cite{Gangopadhyay:2013ofa},

\begin{equation*}
    M_{rem} = \frac{\beta}{4\pi} M_P.
\end{equation*}

Finally, we calculate the entropy of the GUP modified Schwarzschild black hole using the following definition:

\begin{equation*}
    S =\int c^2\frac{dM}{T} = \int C \frac{dT}{T}.
\end{equation*}

Substituting Eq. \ref{T_S} and Eq. \ref{H_C_S}, we obtain the following expression for the entropy.

\begin{equation*}
    S = \frac{k_B}{8\pi}\Bigg[ \frac{1}{2} \left(\frac{M_Pc^2}{k_BT}\right)^2+\beta^2 \ln\left(\frac{k_BT}{M_Pc^2}\right)\Bigg].
\end{equation*}

Using the expression for temperature given in Eq. \ref{T_S}, and performing a binomial expansion, the GUP modified entropy of the Schwarzschild black hole is expressed in terms of the horizon area up to $\mathcal{O}(\beta^2)$, as,

\beq
S = \frac{k_B A_S}{4l_P^2}-\frac{\beta^2}{16 \pi}\Bigg[2+\ln \left(\frac{4\pi A_S}{l_P^2}\right)\Bigg].
\eeq

Here, the first term is the familiar Bekenstein-Hawking entropy for the Schwarzschild black hole with the horizon area $A_S=\pi r_s^2$. The second term represents the leading-order correction to the entropy due to GUP. \\

At this point, one can obtain the GUP modified metric function for the Schwarzschild spacetime that gives the temperature-mass relation as in Eq. \ref{M_Sch}. The required metric function reads \cite{Scardigli:2014qka},

\beq
f_\text{GUP}(r) = 1- \frac{2GM}{c^2 r}+\frac{\lambda G^2 M^2}{c^4 r^2},
\eeq

where the correction parameter ($\lambda$) in the first order approximation is related to $\beta$ by,

\begin{equation*}
    \beta(\lambda) = - \pi^2 \frac{G M^2}{\hbar}\left(\frac{\lambda^2}{1-\lambda}\right).
\end{equation*}

In \cite{Scardigli:2014qka}, the authors observe that the GUP deformation parameter must be negative for the modified Schwarzschild metric to yield the correct GUP-corrected Hawking temperature. Interestingly, this result is consistent with analogous findings in crystal lattice systems, suggesting a possible hint toward a granular or discrete structure of spacetime at the Planck scale \cite{PhysRevD.81.084030}.\\

We now incorporate the Generalized Uncertainty Principle into the black hole solution of five-dimensional Einstein–Gauss–Bonnet gravity. In the following section, we present the resulting modifications to key thermodynamic quantities, including the temperature, heat capacity, and entropy.

\section{General Uncertainty Principle and EGB Black Hole in 5 Dimensions}\label{three}

Building upon the analysis carried out for the Schwarzschild case, we now extend the study to examine the impact of GUP on the thermodynamic behavior of black holes in Einstein–Gauss–Bonnet (EGB) gravity. To begin with, we consider the five-dimensional EGB theory, which is governed by the action,

\begin{equation}
    {\cal A^{GB}} = \frac{1}{16 \pi G_5} \int d^5x \sqrt{-g}\Bigg[R
      +\,\alpha_{GB}\, \left( R^2 -\, 4 \, R_{ab}R^{ab} +\,  R_{abcd} R^{abcd}
      \right)\Bigg].
\end{equation}

Here $\alpha_{GB}$ is the Gauss-Bonnet coupling parameter and $G_5$ denotes Newton's gravitational constant in five dimensions. The theory admits spherically symmetric vacuum black hole solutions of the form \cite{PhysRevLett.55.2656},

\beq
ds^2 = - f(r) dt^2 + \frac{dr^2}{f (r)} + r^2\, d\Omega_3^2.
\eeq

 The metric function $f(r)$ is given by \cite{WILTSHIRE198636,PhysRevD.38.2434,Cai:2001dz,Cai:2013qga},

\beq
f(r)=1+\frac{r^2}{2\alpha}\Bigg[1-\sqrt{1+\frac{32\alpha G_5 M}{3 \pi c^2r^4}}\Bigg], \label{BD}
\eeq

where the parameter $\alpha=\alpha_{GB}/2$. Such black holes are analogous to asymptotically flat, Schwarzschild black holes in general relativity, and the zero of the metric function $f(r)$ determines the location of the horizon ($r_+$). One finds,

\beq \label{mass_radius}
r_+ = \sqrt{\frac{8G_5M}{3 \pi c^2}-\alpha}.
\eeq

Following the steps outlined in the previous section, the position and momentum uncertainties of a particle near the horizon are taken to be,

\begin{equation*}
  \delta x = \epsilon r_+; \quad \delta p = \frac{k_B T}{c}.
\end{equation*}

Now, the GUP relation given by Eq. \ref{GUP} becomes,

\beq \label{GUP_2}
 r_+ = \frac{\hbar c}{2\epsilon}\,\Big[\frac{1}{k_B T}+\beta^2 \frac{k_B T}{\left(M_pc^2\right)^2}\Big].
\eeq

To obtain $\epsilon$, we use the semi-classical expression for the Hawking temperature of a 5D EGB black hole,

\beq \label{temperature}
T = \frac{\hbar c}{4\pi k_B }f'(r_+) = \frac{\hbar c}{2\pi k_B} \frac{r_+}{\left(r_+^2+2\alpha\right)}.
\eeq

When $\beta=0$, the expression for the temperature obtained from Eq. \ref{GUP_2} coincides with the above expression. This condition sets the value of $\epsilon$ as,

\beq
\epsilon = \pi \left(1+\frac{2\alpha}{r_+^2}\right).
\eeq

GUP modified relation between the mass and the temperature of a 5-dimensional EGB black hole is described by,

\beq \label{mass_temp}
r_++\frac{2\alpha}{r_+} =  \frac{\hbar c}{2\pi}\,\Big[\frac{1}{k_B T}+\beta^2 \frac{k_B T}{\left(M_pc^2\right)^2}\Big].
\eeq

Note that when $\alpha = 0$, Eq. \ref{mass_temp}, together with Eq. \ref{mass_radius}, yields:

\beq
M = \frac{3M_p^3 c^4}{32 \pi}\Big[\frac{1}{k_B T}+\beta^2 \frac{k_B T}{\left(M_pc^2\right)^2}\Big]^2.
\eeq

Here, we have used $G_5=\hbar^2/M_p^3$. This is the GUP modified relation between the mass and temperature of a 5-dimensional Schwarzschild-Tangherlini black hole obtained in \cite{Gangopadhyay:2015zma}.\\

In the following section, we derive the GUP modifications to the thermodynamic quantities such as heat capacity and entropy by exploiting Eq. \ref{mass_temp}.

\section{Thermodynamics of GUP-Modified EGB Black hole}\label{four}

The GUP modified temperature can be obtained from Eq. \ref{mass_temp}. Solving the quadratic equation yields two solutions for the temperature, and we choose the one which reduces to Eq. \ref{temperature} in the limit $\beta \to 0$. Thus, the GUP modified temperature of a five-dimensional EGB black hole is given by,

\beq
T = \frac{r_0}{k_B}\left(1-\sqrt{1-\left(\frac{M_pc^2}{r_0 \beta }\right)^2}\right),
\eeq

where,

\begin{equation}\label{r_0}
    r_0 = \frac{\pi (M_pc^2)^2}{\hbar c \beta^2}\frac{\left(r_+^2+2\alpha\right)}{r_+}.
\end{equation}

The critical mass below which the temperature becomes complex is given by,

\beq\label{cri_mass}
M_{cri} = \frac{3M_p \beta^2}{16 \pi}\Bigg[ 1-\frac{\alpha(M_pc^2)^2}{\beta^2}+\sqrt{1-\frac{4\alpha}{\beta^2}(M_pc^2)^2}\Bigg].
\eeq

The heat capacity of the black hole is obtained as,

\begin{eqnarray}\label{heat_cap}
  C = c^2 \frac{dM}{dT}  = \frac{3k_B M_pc}{8 \hbar}\left(\frac{r_+^3}{r_+^2-4\alpha}\right)
    \Bigg[\beta^2-\left(\frac{M_pc^2}{k_BT}\right)^2\Bigg].
\end{eqnarray}

We observe from Eq. \ref{heat_cap} that the heat capacity becomes negative in the regime \( k_B T \ll M_P c^2 \). This implies that, as the black hole radiates and loses mass, its temperature increases. The temperature continues to rise until the heat capacity vanishes, indicating a turning point in the thermodynamic behavior. Beyond this point, the radiation process ceases, and the black hole mass no longer evolves with temperature. The system ultimately settles into a stable configuration characterized by a finite remnant mass and a corresponding finite temperature. The remnant mass is obtained from $C=0$ or $\beta = M_p c^2/(k_BT)$. The expression for remnant mass is given by,

\beq
 M_{rem}=\frac{3M_p \beta^2}{16 \pi}\Bigg[ 1-\frac{\alpha(M_pc^2)^2}{\beta^2}+\sqrt{1-\frac{4\alpha}{\beta^2}(M_pc^2)^2}\Bigg].
\eeq

One can see from Eq. \ref{cri_mass} that the critical mass is equal to the remnant mass, similar to the black hole solutions in GR. Now, the entropy can be obtained and expressed using binomial expansion and keeping terms up to $\mathcal{O}(\beta^2)$ as,

\beq
S=\int \frac{C}{T}\left(\frac{\partial T}{\partial r_+}\right) dr_+ = \frac{k_B (M_pc)^3}{2\hbar^3}\pi^2 r_+^3\left(1+\frac{6\alpha}{r_+^2}\right) - \frac{3k_B M_p c r_+}{8\hbar}\Bigg[ 1-\sqrt{\frac{2\alpha}{r_+^2}}\arctan\left(\sqrt{\frac{r_+^2}{2\alpha}}\right)\Bigg]\beta^2.
\eeq

In terms of the area of the horizon $A=2\pi^2 r_+^3$ and Planck length $l_P=\sqrt{G_5\hbar/c^3}$, we have the entropy of the GUP modified EGB black hole as,

\beq\label{GUP_entropy_EGB}
S = \frac{k_B A}{4l_P^2}\left(1+\frac{6\alpha}{r_+^2}\right)-\frac{3k_Br_+}{8l_P}\Bigg[ 1-\sqrt{\frac{2\alpha}{r_+^2}}\arctan\left(\sqrt{\frac{r_+^2}{2\alpha}}\right)\Bigg]\beta^2.
\eeq

The first term is the familiar expression for the EGB entropy, and the second term corresponds to the correction due to GUP. Note that in the limit $\alpha \to 0$ the second term vanishes, indicating the entropy is modified only at $\mathcal{O}(\beta^4)$. This is consistent with the results presented in \cite{Gangopadhyay:2015zma}.\\

Interestingly, similar to the case of a five-dimensional Schwarzschild black hole, the logarithmic correction to the entropy due to GUP can be recovered if one employs a linear form of GUP, which arises from incorporating DSR into the standard GUP framework. The linear GUP is motivated by Doubly Special Relativity (DSR) theories \cite{Amelino-Camelia:2000stu, Magueijo:2002am}, which extend Einstein's special relativity by postulating, in addition to the invariant speed of light, a second observer-independent scale associated with the Planck energy. The central physical idea behind DSR is that as one approaches the Planck scale, the conventional notions of space and time cease to be classical, and Lorentz transformations must be modified so that all inertial observers agree not only on the speed of light but also on the new fundamental scale. This deformation of relativistic kinematics implies that the energy–momentum relations and the uncertainty relations are modified, leading to a minimal measurable length. In this framework, the commutation relation is modified and the corresponding uncertainty relation becomes $\delta x \delta p \ge \frac{\hbar}{2} (1 - (\gamma l_P\delta p)/\hbar)$. The dimensionless parameter $\gamma$ quantifies the strength of the GUP correction \cite{Ali:2009zq}. Phenomenological studies have constrained the corresponding parameter $\gamma$, with bounds ranging from $10^{10}$ to $10^{21}$ \cite{Scardigli:2014qka, Das:2008kaa, Das:2010zf,  LIGOScientific:2016vbw, Jusufi:2020wmp}, depending on the physical system considered. We treat $\gamma$ as a free parameter within these bounds in our analysis. To this extent, the modified uncertainty relation takes the form:

\begin{equation}\label{newGUP}
    \delta x \delta p \geq \frac{\hbar}{2}\Bigg\{1 - \frac{\gamma l_P}{\hbar}\delta p + \frac{\beta^2 l_P^2}{\hbar^2}(\delta p)^2\Bigg\},
\end{equation}

Following the steps outlined earlier, one obtains the relation between mass and temperature as,

\beq \label{mass-temp}
r_++\frac{2\alpha}{r_+} =  \frac{\hbar c}{2\pi}\,\Big[\frac{1}{k_B T}-\frac{\gamma}{M_Pc^2}+\beta^2 \frac{k_B T}{\left(M_pc^2\right)^2}\Big].
\eeq

The temperature of the black hole in five-dimensional EGB theory due to the general uncertainty relation is obtained by solving the above expression.

\beq
T = \frac{\left(r_0+r_0'\right)}{k_B}\left(1-\sqrt{1-\left(\frac{M_pc^2}{r_0 \beta +r_0'}\right)^2}\right),
\eeq
where $r_0$ is given in Eq. \ref{r_0} and $r_0' = \gamma M_P c^2/(2\beta^2)$. Finally, the entropy is deduced up to $\mathcal{O}(\gamma \beta^2)$ as,

\begin{eqnarray}
S &=& \frac{k_B A}{4l_P^2}\left(1+\frac{6\alpha}{r_+^2}\right)-\frac{3\pi(\pi-2)k_B (M_Pc^2)^2r_+^2}{8\hbar^2}\  \gamma-\frac{3k_Br_+}{8l_P}\Bigg[ 1-\sqrt{\frac{2\alpha}{r_+^2}}\arctan\left(\sqrt{\frac{r_+^2}{2\alpha}}\right)\Bigg]\beta^2\\ \nonumber
&-&\frac{3(3\pi-4)k_B}{16 \pi}\left(\frac{\alpha}{r_+^2+2\alpha}+\frac{1}{2}\ln\Big[r_+^2+2\alpha\Big]\right) \gamma \beta^2 + \mathcal{O}(\gamma^3, \beta^4)
\end{eqnarray}

We immediately notice the emergence of a logarithmic term among the higher-order corrections to the entropy when the DSR-modified GUP is applied to five-dimensional EGB black holes. In the limit $\gamma \to 0$, the entropy expression correctly reduces to the standard GUP-modified entropy of the EGB black hole obtained in Eq. \ref{GUP_entropy_EGB}. Interestingly, this behavior closely resembles what has been previously observed in the case of five-dimensional Schwarzschild black holes \cite{Gangopadhyay:2015zma}. However, a key difference arises: in the Schwarzschild case, the logarithmic correction typically appears as the leading-order quantum correction to the semiclassical entropy, whereas in the EGB scenario, the logarithmic term arises only as a higher-order correction. This distinction underscores the role played by the underlying gravitational theory in shaping the form and order of quantum corrections to black hole entropy. Further, when both modification parameters $\beta$ and $\gamma$ become zero, we obtain the entropy of the five-dimensional Einstein-Gauss-Bonnet black hole as expected.\\

\section{discussion}\label{dis}

In this work, we investigate the thermodynamic modifications of five-dimensional Einstein-Gauss-Bonnet black holes induced by the Generalized Uncertainty Principle. By considering a massless particle in the vicinity of the event horizon, under the assumption of local thermodynamic equilibrium, we derive the GUP-modified mass–temperature relation for the EGB black hole. The thermodynamic stability of the black hole is examined through the behavior of its heat capacity. Our analysis reveals that, contrary to the standard Hawking evaporation scenario, the black hole does not completely radiate away its mass. Instead, it asymptotically approaches a stable remnant configuration with a finite, non-zero temperature, i.e., a phenomenon previously observed in GUP-corrected black hole solutions within general relativity.\\

Furthermore, we derive the GUP-corrected entropy of the five-dimensional EGB black hole up to the leading non-trivial order, $\mathcal{O}(\beta^2)$. We find that this correction term arises purely due to the presence of the Gauss-Bonnet term in the gravitational action and vanishes when the Gauss-Bonnet coupling constant is set to zero. This observation is consistent with the earlier result that, for the five-dimensional Schwarzschild-Tangherlini black hole, the GUP-induced corrections to entropy begin only at $\mathcal{O}(\beta^4)$ \cite{Gangopadhyay:2015zma}.\\

It is widely reported in the literature that logarithmic corrections to black hole entropy are expected to be universal. In particular, small statistical fluctuations around thermodynamic equilibrium typically give rise to logarithmic terms as subleading corrections to the Bekenstein-Hawking entropy. However, in our analysis of the five-dimensional Einstein-Gauss-Bonnet black hole, we find that the correction to entropy induced by the Generalized Uncertainty Principle is non-logarithmic. Interestingly, this non-logarithmic behavior is not unique to the EGB case as it has also been observed in the GUP-corrected entropy of black holes in five- and seven-dimensional Schwarzschild-Tangherlini spacetimes. This suggests that such a feature could be a general outcome in certain higher-dimensional gravitational settings. However, when we incorporate a modified form of GUP derived by embedding Doubly Special Relativity, as defined in Eq. \ref{newGUP}, a logarithmic term reappears in the entropy expression. This result aligns with previous findings for five-dimensional Schwarzschild black holes, where the DSR-modified GUP yields a logarithmic correction. What is particularly notable in the EGB case is that, unlike in General Relativity, where the logarithmic term arises as a leading-order correction, the logarithmic term appears only at sub-leading order. This distinction highlights the influence of the underlying gravity theory on the nature of quantum corrections.\\

Based on these observations, we propose that the GUP-induced corrections to black hole thermodynamics are sensitive not only to the dimensionality of spacetime but also to the choice of gravitational theory. We further conjecture that, similar to the Schwarzschild case in higher dimensions, the emergence of logarithmic corrections in the entropy of odd-dimensional EGB black holes may require the inclusion of DSR modifications in the GUP framework. Exploring this conjecture further by systematically studying GUP-induced entropy corrections across various black hole models and dimensions could shed new light on the interplay between quantum gravity effects, spacetime geometry, and dimensionality. Also, following \cite{Scardigli:2014qka}, one can deduce the modified EGB metric due to GUP corrections and study the thermodynamics of EGB black hole solutions in five or higher dimensions. Another possible extension of our work would be to revisit the calculation of Hawking radiation using the GUP-modified metric of EGB spacetime, incorporating self-interaction effects as presented in \cite{Bhandari:2024xcr, Fairoos:2017piu}. We leave these for future studies.

\section{Acknowledgements}

SKP and CF gratefully acknowledge the hospitality of the Inter-University Centre for Astronomy and Astrophysics (IUCAA), Pune, during the Refresher Course on Astronomy and Astrophysics 2024, where the initial stages of this work were carried out.

\appendix

\bibliography{BibTex}



\end{document}